\documentstyle{ioplppt}

 \jl{3}
\input{epsf}
\begin{document}

\title{Surface Properties of the Half- and Full-Heusler Alloys}

\author{I Galanakis
\ftnote{3}{To
whom correspondence should be addressed, e-mail:
I.Galanakis@fz-juelich.de} }

\address{Institut f\"ur Festk\"orperforschung,
Forschungszentrum J\"ulich, D-52425 J\"ulich, Germany}

\begin{abstract}
Using a full-potential \textit{ab-initio} technique I  study the
electronic and magnetic properties of the (001) surfaces of the
half-Heusler alloys, NiMnSb, CoMnSb and PtMnSb and of the
full-Heusler alloys Co$_2$MnGe, Co$_2$MnSi and Co$_2$CrAl. The
MnSb terminated surfaces of the half-Heusler compounds present
properties similar to the bulk compounds
and, although the half-metallicity is lost,
an important spin-polarisation at the Fermi level.
In contrast to this the Ni terminated surface shows an almost
zero net spin-polarisation. While the  bulk Co$_2$MnGe and
Co$_2$MnSi are almost half-ferromagnetic, their surfaces lose the
half-metallic character and the net spin-polarisation at the Fermi
level is close to zero. Contrary to these compounds the CrAl terminated
(001) surface of Co$_2$CrAl shows a spin polarisation of about
84\%.
\end{abstract}


\maketitle

\section{Introduction}
After the discovery of giant magnetoresistance (GMR) by the groups of Fert \cite{Fert}
and Gr\"unberg \cite{Grunberg} in 1988, a new field  in condensed matter,
the magneto- or spinelectronics, evolved which has grown steadily in the last ten
years. One of the most interesting
problems of this new field is the spin-injection from a
ferromagnet into a semiconductor, that would lead to the creation
of efficient spin-filters \cite{Kilian00}, tunnel junctions
\cite{Tanaka99}, GMR devices for spin injection \cite{Caballero98}, etc. This has
intensified the interest in the so-called half-ferromagnetic
materials which have a band gap at the Fermi level ($E_F$) for one
spin direction and thus exhibit
 100\% spin polarisation at $E_F$.
So in principle during the spin-injection process only spin-up
electrons would be injected in the semiconductor allowing the
creation of the perfect spin-filter and spin-dependent devices with 
superior performances.

Attractive candidates for half-ferromagnetic materials are the 
Heusler alloys of which two distinct families exist. The compounds of the 
first family have the form XYZ and crystallize in
the $C1_b$ structure, which consists of 4 fcc sublattices occupied
by the three atoms X, Y and Z and a vacant site \cite{landolt}.
They are also known as half-Heusler compounds. In 1983 de Groot
and his collaborators \cite{groot} showed that one of them,
NiMnSb, has a gap at $E_F$ in the minority band. Also PtMnSb
\cite{iosif,youn} and  CoMnSb \cite{Kubler84} have been predicted
to be half-ferromagnets. Infrared absorption \cite{Kirillova95}
and spin-polarized positron-annihilation \cite{Hanssen90}
experiments have verified the half-ferromagnetic character of bulk
NiMnSb.  There is also ellipsometric evidence of the spin-down gap
for PtMnSb \cite{Bobo97}.

Recently it has become possible to grow high quality films of
Heusler alloys and it is mainly NiMnSb that has attracted the
attention \cite{Roy}. Unfortunately these films were found not to
be half-ferromagnetic \cite{Soulen98,Mancoff99,Zhu01,Bona}; a maximum
value of 58\% for the spin-polarisation of NiMnSb was obtained by
Soulen \textit{et al.} \cite{Soulen98}. These polarisation values
are consistent with a small perpendicular magnetoresistance
measured for NiMnSb in a spin-valve structure \cite{Caballero99},
a superconducting tunnel junction \cite{Tanaka99} and a tunnel
magnetoresistive junction \cite{Tanaka97}. Ristoiu \textit{et al.}
showed that during the growth of the NiMnSb thin films, Sb and
then Mn atoms segregate to the surface, which is far from being
perfect, thus decreasing the obtained spin-polarisation
\cite{Ristoiu00}. But when they removed the excess of Sb by a
flash annealing, they managed to get a nearly stoichiometric
ordered alloy surface being terminated by a MnSb layer, which
presented a spin-polarisation of about 67$\pm$9\% at room
temperature \cite{Ristoiu00}. The temperature dependence of the
spin moments for such a film was studied by Borca \textit{et al.}
\cite{Borca01}. Wijs and de Groot have shown by first-principle
calculations that NiMnSb surfaces do not present 100\%
spin-polarisation and they proposed that at some interfaces it is
possible to restore the half-ferromagnetic character of NiMnSb
\cite{groot2}. Also recently, Jenkins and King studied by a
pseudopotential technique the MnSb terminated (001) surface of
NiMnSb and showed that there are two surface states at the Fermi
level, which are well localized at the surface layer
\cite{Jenkins01} and they persist even when the MnSb surface is covered by a 
Sb overlayer \cite{Jenkins02}. They found also that the surface only mildly
reconstructs; the Sb atoms move outwards, the Mn atoms inwards
with a total buckling of only 0.06 \AA , and this small $c(1\times
1)$ reconstruction is energetically more favourable than the
creation of Mn or Sb dimers.

The second family of Heusler alloys are the so-called full-Heusler
alloys and they have the X$_2$YZ formula. They crystallize in the $L2_1$
structure which is similar to the $C1_b$ structure but now also
the vacant site is occupied by a X atom. They have attracted a lot
of attention due to the diverse magnetic phenomena they present
\cite{landolt,Pierre97} and mainly the transition from  a
ferromagnetic phase to an antiferromagnetic one by changing the
concentration of the carriers \cite{Kubler83}. Webster
\cite{Webster} was the first to   synthesise full-Heusler alloys
containing Co, and Ishida and collaborators \cite{Ishida95} have
proposed that the compounds of the type Co$_2$MnZ, where Z stands
for Si and Ge, are also half-ferromagnets. But Brown \textit{et
al.} \cite{Brown} using polarized neutron diffraction measurements
have shown that there is a finite very small spin-down DOS at the
Fermi level instead of an absolute gap. Recently, Ambrose
\textit{et al.} managed to grow a Co$_2$MnGe thin film on a GaAs(001)
substrate by molecular beam epitaxy \cite{Ambrose}, and 
Geiersbach \textit{et al.}  grew by sputtering (110) thin films of 
Co$_2$MnSi, Co$_2$MnGe and Co$_2$MnSn using a metallic seed on top of a
MgO(001) substrate \cite{Geiersbach}.

In this communication I study the (001) surfaces of the
NiMnSb, CoMnSb and PtMnSb half-ferromagnetic materials taking into
account the two different possible terminations. I compare their
magnetic and electronic properties with the bulk calculations and
can explain the large spin-polarisation obtained for the MnSb
terminated (001) surface of NiMnSb. The second part of my  study
is devoted to the (001) surfaces of the Co$_2$MnGe and Co$_2$CrAl
compounds. The substitution of Mn by Cr leads to a change in the
electronic properties of the studied surfaces and the CrAl
terminated surfaces shows a very large spin-polarisation of about
84\%.

\section{Method}

\begin{figure}
   \begin{center}\begin{minipage}{2.5in}
  \epsfxsize=3.0in \epsfysize=1.7in \centerline{\epsfbox{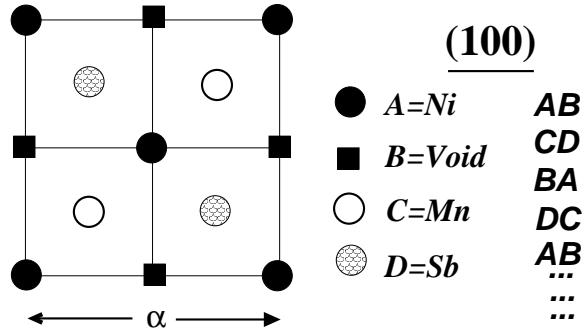}}
  \end{minipage} \end{center}
\caption{\label{ios1} Schematical representation of the (001)
surface of NiMnSb. There are two possible
 terminations: i) MnSb and ii) Ni-Void. The interlayer distance is
0.25$a$. In the $i\pm 2$ layer the atoms have
 exchanged positions compared to the $i$ layer. In the case of the 
Co$_2$MnGe compound the Co atoms
 occupy the Ni and Void sites.}
\end{figure}

In the calculations I used the full-potential version of the screened
Korringa-Kohn-Rostoker (KKR) Green's function method
\cite{zeller95,Pap02} in conjunction with the local
spin-density approximation \cite{vosko}. The full-potential is
implemented by using a Voronoi construction  of Wigner-Seitz
polyhedra that fill the space \cite{Pap02}. I  have used a
two-dimensional 30$\times$30 {\bf q}$_\parallel$-space grid to
perform the integrations in the first surface Brillouin zone. To evaluate
the charge density one has to integrate the Green's function over an
energy contour in the complex energy plane; for this 42 energy
points were needed. For the wavefunctions I took angular momentum up to
 $\ell_{max}=3$ into account  and for the
charge density up to $\ell_{max}=6$. Finally a
tight-binding cluster of 51 atoms was used in the calculation of
the screened KKR structure constants \cite{zeller97}. To simulate
the surface I used a slab with 15 metal layers embedded in
half-infinite vacuum from each side. Such a slab has two
equivalent surfaces avoiding the creation of slab-dipoles. This
slab thickness is enough so that the layers in the middle exhibit
bulk properties; they show a spin-down gap of the same width as in
the bulk and the same relative position of the Fermi level and
finally the magnetic moments differ less than 0.01$\mu_B$ from the
bulk values. In figure \ref{ios1} I  present the structure of the
(001) surfaces in the case of NiMnSb. There are two different
possible terminations, one containing the Mn and Sb atoms while the
other contains the Ni atom and the vacant site. In the case of the
full-Heusler alloys like Co$_2$MnGe, there are also two possible
surface terminations: the first one containing the Mn and Ge atoms
in the surface layer while the second one is Co terminated (in figure~\ref{ios1} both
the Void and Ni sites are  occupied by  Co atoms). The interlayer
distance is 0.25 the lattice constant. I  have used in all my 
calculations the experimental lattice constants \cite{landolt}. In
the perpendicular direction the layer occupancy is repeated every
fourth layer, since in the $i\pm 2$ layer the atoms have exchanged
positions compared to the $i$ layer.

\section{Half-Heusler Alloys}

\subsection{Density of states}

\begin{figure}
   \begin{center}\begin{minipage}{3.0in}
\epsfxsize=2.8in \epsfysize=2.8in \centerline{\epsfbox{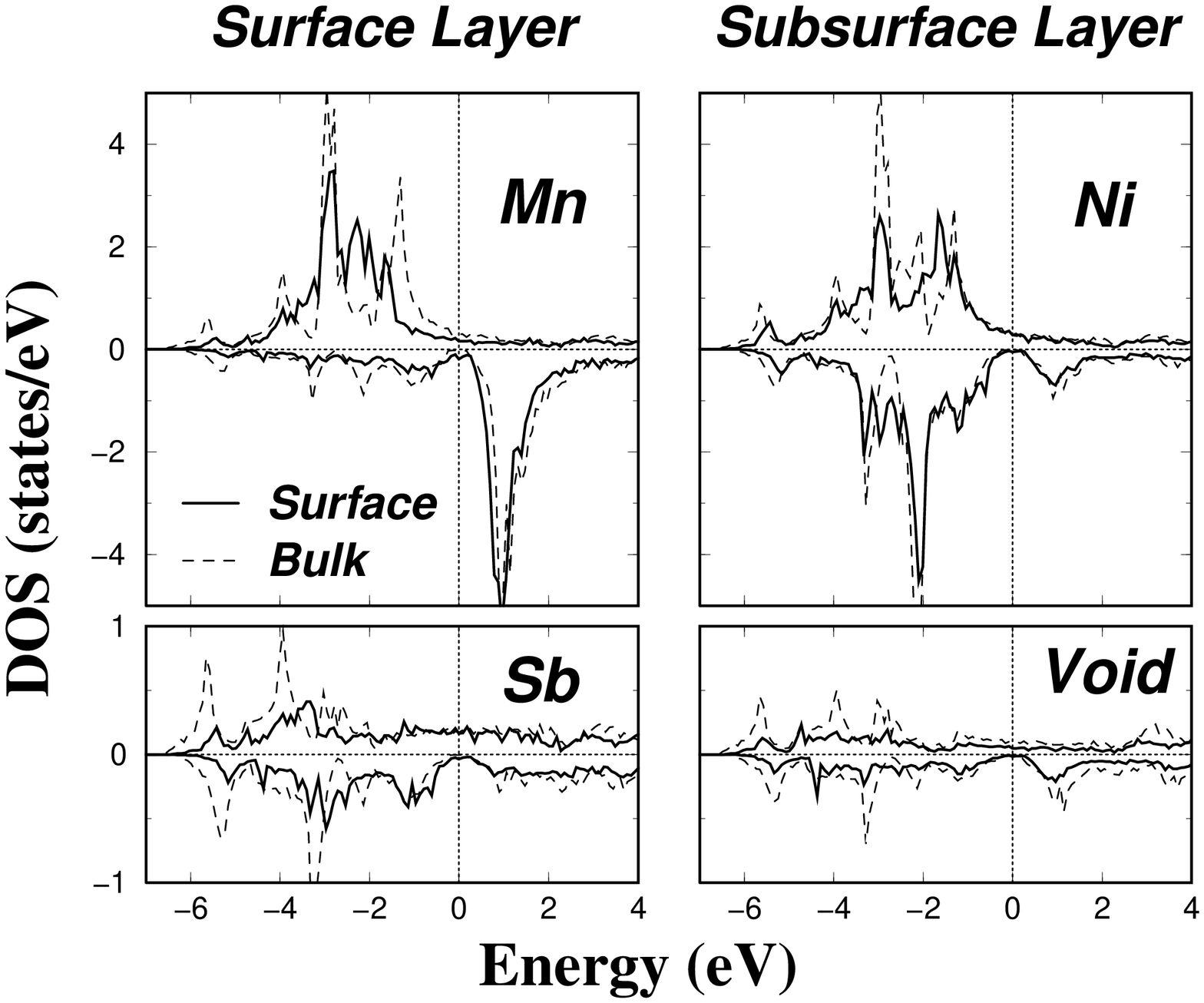}}
\end{minipage}\begin{minipage}{3.0in}
\epsfxsize=2.8in \epsfysize=2.8in \centerline{\epsfbox{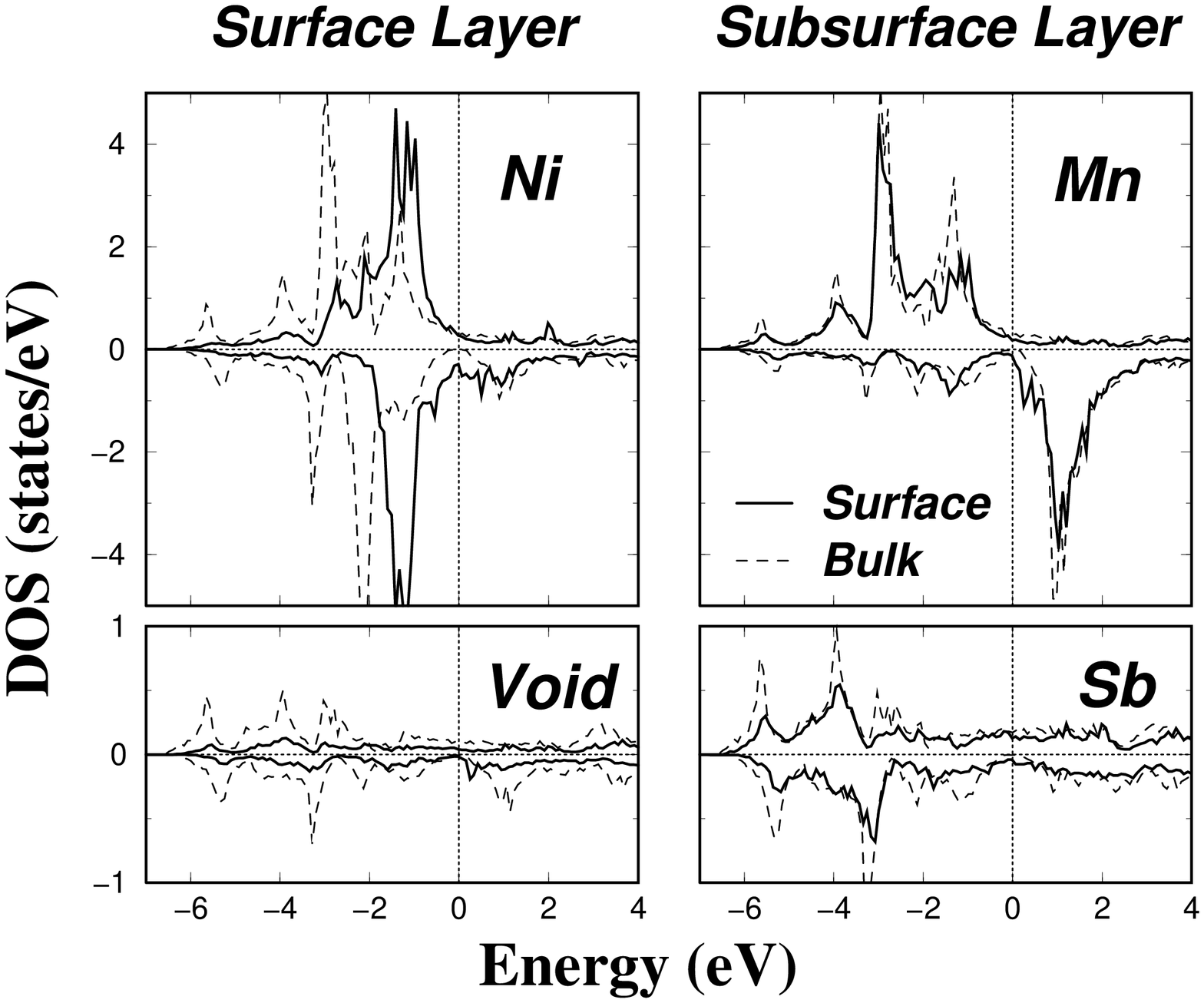}}
\end{minipage} \end{center}
\caption{\label{ios2} Spin- and atom-projected DOS for the
MnSb-terminated NiMnSb(001) surface  
(left panel) and for  the Ni-Void terminated (001) surface
(right panel). The dashed lines give the local DOS of the atoms in
the bulk.}
\end{figure}

In the first part of my  study I concentrate on the
half-Heusler compounds and more specifically on the (001) surfaces
of the NiMnSb, CoMnSb and PtMnSb. I  use the NiMnSb compound as
the model system   and at the end of the section I   discuss its differences
with the other two alloys. As already discussed in the
introduction, the MnSb terminated surface of NiMnSb shows a very
small reconstruction, while no-information is available for the
Ni-terminated surface which in principle should show a large
reconstruction due to the vacant site at the surface. In my  study
I  assume an ``ideal'' epitaxy in both cases. In the left panel of
figure \ref{ios2} I  present the atom- and spin-projected density
of states (DOS) for the Mn and Sb atoms in the surface layer and
the Ni and vacant site in the subsurface layer for the MnSb
terminated NiMnSb(001) surface, and in the right panel of the same
figure I  present the atom- and spin-projected DOS of
the surface and the subsurface layers for the Ni
terminated surface. In both cases I  compare the surface DOS with
the bulk calculations (dashed line) \cite{iosif01}. 

In the case of
the MnSb terminated surface the DOS with the exception of the gap
area is very similar to the bulk calculations. The Ni atom in the
subsurface layer presents practically a half-ferromagnetic
character with an almost zero spin-down DOS, while for the bulk there is 
an absolute gap. The spin-down band of the vacant site
also presents a very small DOS around the Fermi level. The Mn and
Sb atoms in the surface layer show  more pronounced differences
with respect to the bulk, and  within the gap  there is a very small
Mn-$d$ and Sb-$p$ DOS. These states are strongly localized  at the
surface layer as at the subsurface layer there is practically no
states inside the gap. This is in agreement with previous
pseudopotentials calculations that showed that the surface states
at the case of the MnSb-terminated NiMnSb(001) surface are
localized at the surface layer \cite{Jenkins01}. Our results are
in agreement with the experiments of Ristoiu \textit{et al.}
\cite{Ristoiu00} who  in the case of a MnSb well ordered (001)
surface measured a  high spin-polarisation.

\subsection{Magnetic moments}

In table~\ref{table1} I  have gathered  the spin magnetic moments
of the atoms in the surface and subsurface layers.  In the case of
the MnSb terminated NiMnSb(001) surface, the Ni and vacant site at
the subsurface layer gain a similar charge as in the case of the
bulk, $\sim$0.5$e^-$ for Ni and $\sim$1.3$e^-$ for the vacant
site. Also the Ni and Void spin moments are comparable to the
bulk situation. The  Mn in the surface layer  loses $\sim$0.3$e^-$
more than the bulk Mn and the Sb atom loses also $\sim$0.1$e^-$
more, due to the spilling out of charge into  the vacuum. 
The spin magnetic moment of the Mn atom in the surface layer increases
with respect to the bulk and reaches the $\sim$4$\mu_B$. This behaviour arises 
from the reduced symmetry of the
Mn atom in the surface which loses two of the four neighbouring Co atoms.
In the majority band this leads to a narrowing of the $d$-DOS and a slight
increase of the $d$ count by 0.10 $e^-$ due to rehybridisation, while in the 
minority valence band the Mn $d$ contribution is decreases by 0.20  $e^-$.
Moreover the splitting between the unoccupied Mn states above $E_F$ 
and the center of the occupied Mn states decreases and at $E_F$ a surface states
appears. I  should also mention here that in the case of
a half-ferromagnetic material the total spin magnetic moment per
unit cell should be an integer since the number of both the total
valence electrons and the spin-down occupied states are integers;
the spin moment in $\mu_B$ is simply the number of uncompensated
spins. The total spin moment for the NiMnSb compound is 4 $\mu_B$
(note that in the KKR method one can get the correct charge only if
Lloyd's formula is used for the evaluation of the charge density.
Otherwise, as it is the case in these calculations, the finite
$\ell$ cut-off results in small numerical inaccuracies). Thus in the
case of the surfaces the half-ferromagnetic character is lost
and an increase of the total spin moment is observed, which is
no more an integer number. 

In the case of the Ni terminated surface, the changes in the DOS
compared to the bulk are more pronounced. The Ni atom in the
surface instead of gaining $\sim$0.5$e^-$ as in the bulk now loses
$\sim$0.15$e^-$. Also the vacant site gains only half of the charge as
in the bulk. Due to charge neutrality the Ni $d$ bands move higher
in energy.  As was the case for the Mn surface atom in the MnSb
terminated surface, the Ni spin magnetic moment is
increased. (see table \ref{table1}). The Mn and Sb atoms in the
subsurface layer present a charge transfer comparable to the bulk
compound and also a comparable spin moment. Within the gap there
is  a small Sb and Mn DOS of comparable intensity to the one of the Mn and Sb atoms at
the surface layer of the MnSb terminated surface. Deeper than the
the third layer from the surface the atoms regain a bulk like
behaviour. The origin of the surface states is not the same for the
two different terminations. In the case of the Ni terminated
surface there is a practically rigid shift (although also the shape of
the peaks and their intensity change) towards higher energies of
the Ni bands and now the Fermi crosses a region of high DOS for
both spin-directions. 
In the case of the MnSb terminated
surface, the DOS of the Mn atom at the surface is very similar to
the bulk case. The small spin-down DOS inside the gap comes from a
$d$ like atomic state of Mn that is shifted with respect to the
continuum due to the lower symmetry of the surface. This peak is
broadened by the interaction of the Mn atom with its neighbours,
forming finally a surface band within the gap. This state is  not
localised only at the Mn atom but extends also to the Sb
neighbouring atoms as electrons are delocalised in the two
dimensions. The Ni atoms at the subsurface are slightly polarized
by the Mn atoms and show a very small DOS within the gap.

\subsection{Spin polarisation}

As  discussed in the previous paragraph, the main difference in
the surface DOS between the two terminations is the different 
origin of the surface states.
This behaviour is also reflected in the spin-polarisation of the
surfaces. In table \ref{table2} I  have gathered the number of
spin-up and spin-down states at the Fermi level for each atom at
the surface and the subsurface layer for both terminations. I 
calculated the spin-polarisation as the ratio between the number
of spin-up states minus the number of spin-down states 
over the total DOS at the Fermi level. $P_1$
corresponds to the spin-polarisation when I  take into account
only the surface layer and $P_2$ when I  also include the
subsurface layer. $P_2$ represents quite well the experimental
situation as the spin-polarisation in the case of films is usually
measured by inverse photoemission which probes only the surface of
the sample \cite{Borca02}. In all cases the inclusion of the subsurface layer
increased the spin-polarisation. In the case of the Ni terminated
surface, the spin-up DOS
at the Fermi level is equal to the spin-down DOS and the net
polarisation $P_2$ is zero. In the case of the MnSb terminated surface
the spin-polarisation increases and now $P_2$ reaches a value of
38\%, which means that the spin-up DOS at the Fermi level is about
two times the spin-down DOS. As can be seen in table \ref{table2}
the main difference between the two different terminations is the
contribution of the Ni spin-down states. In the case of the MnSb
surface the Ni in the subsurface layer has a spin-down DOS at the
Fermi level of 0.05 states/eV, while in the case of the
Ni-terminated surface the Ni spin-down DOS at the Fermi level is
0.40 states/eV decreasing considerably the spin-polarisation for
the Ni terminated surface; the Ni spin-up DOS is the same for both
terminations. It is interesting also to note that for both surfaces
the net Mn spin-polarisation is close  to zero while Sb atoms in
both cases show a large spin-polarisation and the number of the Sb
spin-up states is similar to the number of Mn spin-up states, thus
Sb and not Mn is responsible for the large spin-polarisation of the
MnSb layer in both surface terminations. The calculated $P_2$
value of 38\% for the MnSb terminated surface is smaller than
the experimental value of 67\% obtained by Ristoiu and
collaborators \cite{Ristoiu00} for a thin-film terminated in a
MnSb stoichiometric alloy surface layer. But experimentally no exact details of 
the structure of the film are known and the comparison between experiment and 
theory is not straightforward.

\subsection{CoMnSb(001) and PtMnSb(001)}

\begin{figure}
   \begin{center}\begin{minipage}{2.5in}
  \epsfxsize=3.0in \epsfysize=2.5in \centerline{\epsfbox{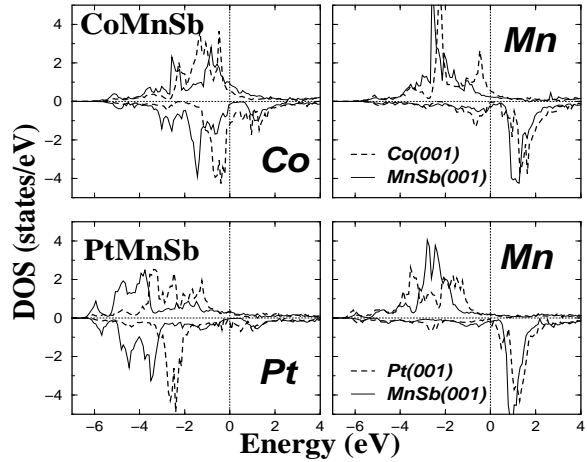}}
\end{minipage} \end{center}
 \caption{\label{ios3} In the upper panel the spin-resolved DOS for Co in the surface
layer  and Mn in the subsurface layer in the case of the
Co-terminated CoMnSb(001) surface and for Co in the subsurface
layer and Mn in the surface layer for a MnSb-terminated
CoMnSb(001) surface. In the bottom panel the spin-resolved DOS of the
 PtMnSb surface for the Pt and MnSb terminations are given.}
\end{figure}

In figure \ref{ios3} I  present the spin-resolved  DOS for the Co
and Mn atoms in the case of the two different terminated
CoMnSb(001) surfaces and the Pt and Mn DOS for the PtMnSb surfaces.
Both CoMnSb and PtMnSb present a behaviour similar to the NiMnSb
surfaces. In the case of the MnSb terminated surfaces the DOS of
the Co or Pt atom in the subsurface layer is similar to the bulk
one and so does the charge transfer. The Mn and Sb atoms at the
surface as was the case for the MnSb-terminated NiMnSb(001)
surface show a small DOS within the gap and these states are
localized at the surface layer. The atom-resolved spin moments are
also close to the bulk values (see table \ref{table1}) but the
total spin magnetic moment is no more an integer reflecting the
loss of the half-metallicity for the surfaces. 
As was the case for the Ni terminated surface the spin magnetic moments of 
the Co and Pt atoms in the case of the
Pt or Co terminated surfaces, respectively, increase considerably with respect to the 
bulk. The Mn and Sb atoms
at the subsurface layer are close to the bulk behaviour exactly as
it was the case for the Ni terminated surface. It is also
interesting to examine the spin-polarisation at the Fermi level. In
the case of the Co terminated CoMnSb surface as it was the case
for the Ni terminated surface there is a shift of the Co spin-down
DOS towards higher energies and the Co spin-down DOS at the Fermi
level is very high and thus $P_2$ is negative, meaning that the
spin-down DOS is larger than the spin-up DOS at the Fermi level.
In the case of the MnSb terminated CoMnSb surface, the Co atom in the
subsurface layer has practically zero spin-down DOS 
and $P_2$ reaches 46\%. In the
case of PtMnSb, Pt does not show such a pronounced difference
between the two surface terminations as the Co atom because it has
practically all its $d$ states filled and the DOS near the Fermi
level is small. But also for PtMnSb the MnSb terminated surface
shows a very large spin-polarisation comparable to the one of       
CoMnSb, while the Pt terminated (001) surface shows a positive
spin-polarisation contrary to the vanishing  net spin-polarisation of
the Ni surface and the negative one  of the Co terminated surface.

\section{Full-Heusler Alloys}

In the second part of my  study I concentrate on the surface
properties of the full-Heusler alloys containing Co. I  have
calculated the Co$_2$MnGe, Co$_2$MnSi and Co$_2$CrAl compounds. Ge and Si are
isoelectronic elements and thus Co$_2$MnGe and Co$_2$MnSi present
similar properties both as bulk systems and as surfaces (they
present similar magnetic moments as can be seen in table
\ref{table3} and similar DOS) and therefore I only discuss
the properties of Co$_2$MnGe. The interest on Co$_2$MnGe arises
mainly for the fact that it is the only full-Heusler alloy that
has been grown on a semiconductor \cite{Ambrose}. Co$_2$MnSi has
the same experimental lattice constant with GaAs and AlAs
\cite{landolt} and Co$_2$CrAl presents a very large spin-up DOS at
the Fermi level due to the smaller exchange-splitting of the Cr-$d$
states compared to the Mn-$d$ states. Our calculations show that
the bulk compounds are not really half-ferromagnets contrary to
the early calculations of Ishida \textit{et al.} using the
linear-muffin tin orbitals  (LMTO) method in the atomic sphere
approximation \cite{Ishida95} but the Fermi level falls within a
broad region of finite very small spin-down DOS. 
The highest occupied bands and the lowest unoccupied bands overlap
slightly destroying the indirect gap \cite{iosif2}. These  results
agree with the experimental results of Brown and collaborators
\cite{Brown} who  using polarized neutron diffraction measurements
have shown that there is a finite, but very small, DOS at the Fermi
level instead of an absolute gap.

\subsection{Co terminated surfaces}

\begin{figure}
   \begin{center}\begin{minipage}{2.55in}
  \epsfxsize=2.5in \epsfysize=2.5in \centerline{\epsfbox{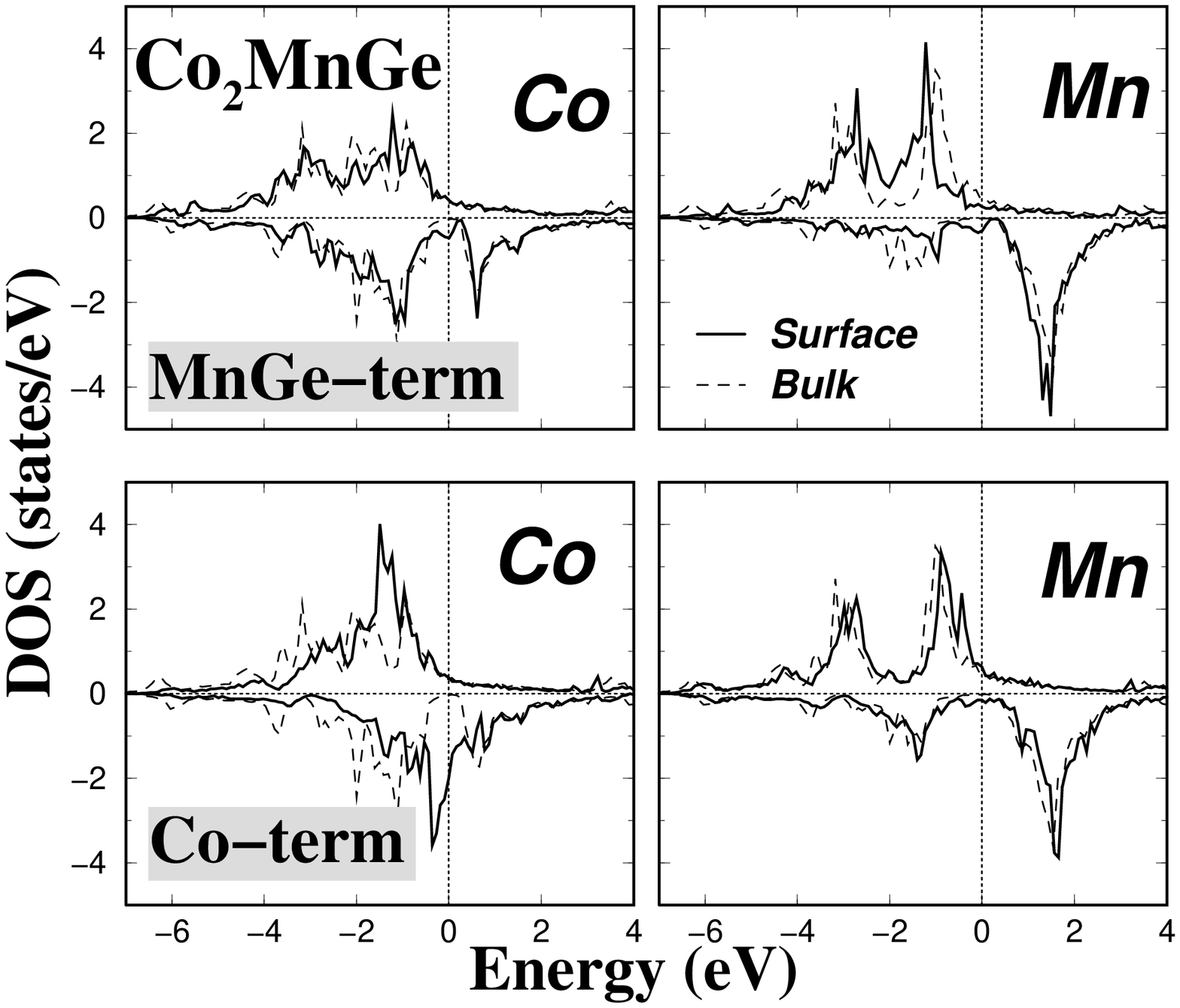}}
\end{minipage}  \begin{minipage}{2.55in}
  \epsfxsize=2.5in \epsfysize=2.5in \centerline{\epsfbox{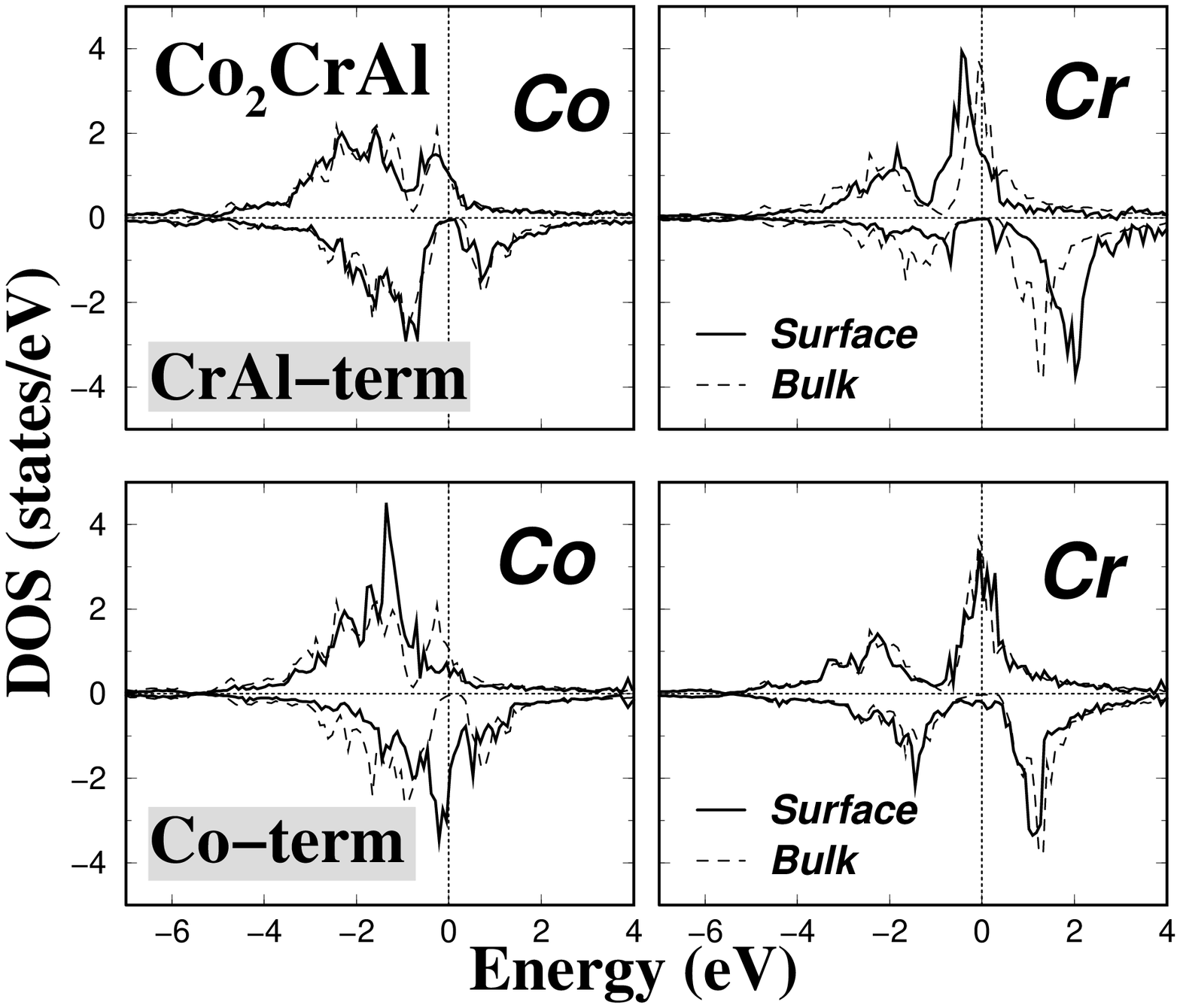}}
\end{minipage} \end{center}
 \caption{\label{ios4} Atom- and spin-projected DOS for the Mn 
atom in the surface and
 the Co atom in the subsurface layer in the case of the MnGe 
terminated Co$_2$MnGe (001) surface
 (left upper panel) and the Co atom in the surface layer and 
the Mn atom in the subsurface for the Co terminated surface
 (left bottom panel). In the right panel similar DOS for the 
Co$_2$CrAl compound. With dashed line the bulk results.}
\end{figure}

In table \ref{table3} I  present the spin-magnetic moments for the
full-Heusler alloys under study and in figure \ref{ios4} the Co and
Mn(Cr) DOS for the two possible terminations for the Co$_2$MnGe
(left panel) and Co$_2$CrAl (right panel) compounds. I  compare in
all cases the surface DOS with the bulk DOS (dashed lines). In the
case of the Co terminated surfaces both compounds show the same
behaviour which is similar to the behaviour of the Co surface atom
in the case of the Co-terminated CoMnSb surface. The lower
coordination number of the Co atoms in the surface layer results
in smaller covalent hybridisation between the Co spin-down $d$
states and the Mn ones  and thus  there is a  practically rigid
shift of the spin-down Co $d$ bands towards higher energies, and
now the Fermi level falls at the edge of the large peak of the minority
spin DOS. Due to this large peak it is very
likely that this surface would reconstruct. 
The Mn and Cr atoms in the
subsurface layer show now a considerable spin-down DOS within the
pseudogap due to the hybridisation with the spin-down $d$ states
of the surface Co atoms. As was the case for the CoMnSb, the spin
moment of the Co at the surface increases to about 1.4 $\mu_B$ per
Co atom while the spin moments of the Mn and Cr atoms in the
subsurface layer slightly decrease due to the larger spin-down DOS
around the Fermi level. Also if one adds the spin-moments for the
surface and subsurface layers, the total spin magnetic moments are no more an almost integer
moment due to the loss of the nearly half-metallicity. Ambrose \textit{et al.} measured a spin 
magnetic moment of 5.1 $\mu_B$ for a Co$_2$MnGe thin film \cite{Ambrose}. This value is larger than
the bulk in agreement with our results but there is no experimental information
on the characterisation of the surface of the sample.

\subsection{MnGe and CrAl terminated surfaces}

In the case of the MnGe terminated Co$_2$MnGe (001) surface the
behaviour is similar to the MnSb terminated surface in the CoMnSb
compound. Due to the reduced symmetry (Mn loses two out of the
four Co nearest neighbours) the hybridisation between the Mn
minority $d$ states and the Co ones is reduced, leading to an increase
of the Mn spin-moment by about 0.6 $\mu_B$, while the Co atom at
the subsurface layer behaves similar to the bulk case. Where the Mn
pseudogap was located in the bulk, there is now a small peak  due to a $d$
like Mn atomic state that is shifted in energy due to the lower
symmetry and which is pinned at the Fermi level. This state
although located at the surface layer is not well localised and the
Co atoms in the subsurface layer present a similar peak at the
Fermi level. Ishida \textit{et al.} \cite{Ishida98} have studied
the MnSi and MnGe terminated Co$_2$MnSi and Co$_2$MnGe surfaces
using a 13 layers thick film. They claim that in the case of the
MnSi surface the half-ferromagnetic character, which they have
calculated for the bulk Co$_2$MnSi \cite{Ishida95}, is preserved,
while in the case of the MnGe surface, surface states
destroy the gap (in the paper they present the results for
the MnSi surface and only shortly refer to the MnGe surface).
These results are peculiar since they get a similar electronic
structure for the bulk compounds and there is no obvious reason
obliging only the MnGe surface to present surface states. Mn atoms
have the same environment in both cases and the hybridisation
between Mn and the $sp$ atoms is similar for both Ge and Si. A
plausible reason for this behaviour is the use of the atomic-sphere
approximation in their calculations, where the potential and the
charge density are supposed to be spherically symmetric. Although this
approximation can accurately describe the bulk compounds
due to the close-packed structure they adopt, it is not suitable
for surfaces where the non-spherical contributions to the
potential and the charge density are important.

The case of Co$_2$CrAl is different from Co$_2$MnGe.  
In line with the reduction of the total
valence electrons by 2, the Cr moment is rather small (1.54 $\mu_B$)
yielding a total moment of only 3 $\mu_B$ instead of 5 $\mu_B$ for Co$_2$MnGe.
The Co terminated Co$_2$CrAl(001) surface shows a similar behaviour 
as the corresponding surface of Co$_2$MnGe , being in both cases dominated 
by a strong Co peak in the gap region of the minority band. However the CrAl 
terminated Co$_2$CrAl surface behaves very differently, being driven by the large
surface enhancement of the Cr moment from 1.54 $\mu_B$ to 3.12 $\mu_B$. As a
consequence the splitting of the Cr peaks in the majority and
minority bands is even enlarged and in particular in the minority band 
the pseudogap is preserved. Thus this surface is a rare case, since for all
the other surfaces studied in this paper, the half-metallicity is destroyed by
surface states.

\subsection{Spin polarisation}

In the last part of my  study I  will discuss the
spin-polarisation for the surfaces of the full-Heusler alloys. I 
will concentrate on the MnGe and CrAl terminated surfaces as the
Co-terminated ones might show large reconstructions and
thus are not interesting for applications. In table \ref{table4}
I  present the atom-resolved spin-up and spin-down DOS at the
Fermi level and the obtained spin-polarisation.
In the case of the MnGe terminated surface the surface states
completely kill the spin polarisation as the majority spin DOS
is pretty  small. In the case of the CrAl terminated surface the
situation is completely different. The minority DOS around the
Fermi level is the same for both the bulk and the CrAl surface.
The Fermi level falls within a region of very high Cr and Co majority
spin DOS as can be seen in table \ref{table4} and thus 92\% of the
electrons at the Fermi level are of spin-up character and $P_2$ reaches 
84\%.

\section{Summary}

I have performed \textit{ab-initio} calculations based on the full-potential
version of the screened KKR Green's function method for the 
(001) surfaces of a series of half-ferromagnetic Heusler alloys, \textit{i.e.}
for the half-Heuslers NiMnSb, CoMnSb and PtMnSb with the $C1_b$ structure and
the full-Heuslers Co$_2$MnGe, Co$_2$MnSi and Co$_2$CrAl with the 
$L2_1$ structure. For the half-Heusler alloys, the MnSb terminated (001) surfaces
 present electronic and
magnetic properties similar to the bulk compounds. There is
however a small finite Mn-$d$ and Sb-$p$ DOS within the bulk spin-down gap and
these surface states are strongly localized at the surface layer.
The spin-polarisation at the Fermi level for this termination reaches the 
38\%. The (001) surfaces terminated at Ni,
Co or Pt present a quite large density of states at the Fermi
level and properties considerably different from the bulk and the
MnSb terminated surfaces. In the case of the Co-based full-Heusler
alloys, the Co terminated surfaces show a behaviour similar to the
Co terminated CoMnSb surface and the Co minority states shift
practically in a rigid way towards higher energies, destroying the
pseudogap. In the case of the MnGe terminated Co$_2$MnGe(001)
surfaces,  the surface states kill completely the spin
polarisation but in the case of CrAl the combination of the very
high majority-spin Cr and Co DOS and the absence of surface states
within the pseudogap result in a very high spin-polarisation of
around 84\%. Thus of all the investigated surfaces, this is the only one 
which preserves the nearly half-metallicity at the surface.

\ack{The author acknowledges financial support from the RT Network
of {\em Computational Magnetoelectronics} (contract
RTN1-1999-00145) of the European Commission. The author would like also
to thank professor P. H. Dederichs for helpful discussions and for a critical reading of 
the manuscript.}


\newpage

\begin{table}
\caption{Spin moments in $\mu_B$ for the Ni-, Pt- and CoMnSb
compounds in the case of i) bulk compounds; ii) the Mn and Sb
atoms in the surface and the Ni (Pt, Co) and the vacant site in
the subsurface layer for the MnSb-terminated (001) surfaces; iii)
like (ii) for the Ni, Pt or Co terminated surfaces. For the surfaces 
the ``total'' moment denotes the sum of the moments in the surface and 
subsurface layers.}
  \label{table1} \begin{indented}
 \item[]
\begin{tabular}{rlrrrrr} \br
\multicolumn{2}{c}{$m^{spin}(\mu_B)$} & Ni (Pt, Co) & Mn & Sb &
Void & Total \\ \mr  NiMnSb & Bulk      & 0.26 & 3.70 & -0.06 &
0.05 & 3.96\\
       & (001)MnSb & 0.22 & 4.02 & -0.10 & 0.04 &4.19\\
       & (001)Ni   & 0.46 & 3.84 & -0.05 & 0.05 &4.30\\
PtMnSb & Bulk      & 0.09 & 3.89 & -0.08 & 0.04 &3.94\\
       & (001)MnSb & 0.08 & 4.19 & -0.13 & 0.03 &4.17\\
       & (001)Pt   & 0.27 & 4.15 & -0.04 & 0.04 &4.42\\
CoMnSb & Bulk      &-0.13 & 3.18 & -0.10 & 0.01 &2.96\\
       & (001)MnSb &-0.06 & 3.83 & -0.12 & 0.01 &3.65\\
       & (001)Co   & 1.19 & 3.31 & -0.09 & 0.02 &4.43 \\ \br
\end{tabular}
\end{indented}
\end{table}

\begin{table}
\caption{Atomic-resolved spin-up and spin-down DOS at the Fermi
level in states/eV units. They are presented as ratios spin-up
over spin-down. Polarization ratios at the Fermi level are
calculated taking into account only the surface layer $P_1$, and
both the surface and subsurface layers $P_{2}$.}
  \label{table2} \begin{indented}
 \item[]
\begin{tabular}{rccccrr} \br
& \multicolumn{4}{c}{MnSb-termination}& & \\ \mr &
\multicolumn{2}{c}{Surface Layer} &\multicolumn{2}{c}{Subsurface
Layer}&&\\ & Mn ($\uparrow/ \downarrow$) & Sb ($\uparrow/
\downarrow$) & Ni[Co,Pt] ($\uparrow / \downarrow$) & Void
($\uparrow / \downarrow$) & $P_1$  ($\uparrow - \downarrow \over \uparrow +
\downarrow$) & $P_{2}$ ($\uparrow - \downarrow \over \uparrow + \downarrow$)
\\ NiMnSb & 0.16/0.19 &0.17/0.03 & 0.28/0.05 & 0.05/0.02& 26\% &
38\% \\ CoMnSb & 0.23/0.27 &0.16/0.07 & 0.91/0.15 & 0.07/0.02 &
6\% & 46\% \\ PtMnSb & 0.21/0.24 &0.31/0.06 & 0.38/0.04 &
0.08/0.02 & 26\% & 46\% \\ &&&&&&\\ \mr &
\multicolumn{4}{c}{Ni(Co,Pt)Void-termination}&& \\ \mr &
\multicolumn{2}{c}{Subsurface Layer} &\multicolumn{2}{c}{Surface
Layer}&&\\ & Mn ($\uparrow / \downarrow$) & Sb ($\uparrow /
\downarrow$) &
 Ni[Co,Pt] ($\uparrow/ \downarrow$) & Void ($\uparrow/ \downarrow$) &
$P_1$  ($\uparrow - \downarrow \over \uparrow + \downarrow$)  &  $P_{2}$
($\uparrow - \downarrow \over \uparrow + \downarrow$) \\ NiMnSb & 0.18/0.16 &
0.13/0.05&  0.27/0.40 &0.04/0.02 & $-$16\% & 0\% \\ CoMnSb &
0.55/0.68 & 0.12/0.07&  0.54/1.15 &0.05/0.04 & $-$34\% & $-$22\% \\
PtMnSb & 0.18/0.14 & 0.21/0.07&  0.30/0.24 &0.05/0.02 & 14\% &
22\% \\ \br
\end{tabular}
\end{indented}
\end{table}

\begin{table}
\caption{Spin moments in $\mu_B$ for the Co$_2$MnGe, Co$_2$MnSi
and Co$_2$CrAl compounds in the case of i) bulk compounds; ii) the
Mn(Cr) and Ge(Si,Al) atoms in the surface and the Co atoms in the
subsurface layer for the MnGe(MnSi,CrAl)-terminated (001)
surfaces; iii) like (ii) for the Co terminated surfaces.  For the surfaces 
the ``total'' moment denotes the sum of the moments in the surface and 
subsurface layers.}
  \label{table3} \begin{indented}
 \item[]
\begin{tabular}{rlrrrr} \br
\multicolumn{2}{c}{$m^{spin}(\mu_B)$} & Co & Mn(Cr) & Ge(Si,Al) &
Total \\ \mr
 Co$_2$MnGe & Bulk & 0.98 & 3.04 & -0.06 & 4.94\\
           & (001)MnGe & 0.96 & 3.65 & -0.10 & 5.47\\
           & (001)Co   & 1.40 & 2.85 & -0.09 & 5.56 \\
Co$_2$MnSi & Bulk      & 1.02 & 2.97 & -0.07 & 4.94\\
           & (001)MnSi & 0.95 & 3.56 & -0.13 & 5.33\\
           & (001)Co   & 1.29 & 2.72 & -0.11 & 5.19 \\
Co$_2$CrAl & Bulk      & 0.76 & 1.54 & -0.10 & 2.96 \\
           & (001)CrAl & 0.76 & 3.12 & -0.02 & 4.62\\
           & (001)Co   & 1.36 & 1.06 & -0.11 & 3.67 \\ \br
\end{tabular}
\end{indented}
\end{table}

\begin{table}
\caption{Same as Table~\protect{\ref{table2}} for the MnGe and
CrAl terminated (001) surfaces of the Co$_2$MnGe and
 Co$_2$CrAl compounds, respectively.}
  \label{table4} \begin{indented}
 \item[]
\begin{tabular}{rrr} \br
 \multicolumn{3}{c}{MnGe and CrAl terminations} \\ \mr
  &Co$_2$MnGe& Co$_2$CrAl \\ Mn-Cr ($\uparrow/ \downarrow$) &
0.26/0.30 & 1.48/0.03 \\ Ge-Al ($\uparrow/ \downarrow$) &
0.22/0.18 & 0.01/0.15 \\ Co    ($\uparrow/ \downarrow$) &
0.41/0.48 & 1.03/0.06 \\ $P_1$  ($\uparrow - \downarrow \over \uparrow +
\downarrow$) & 0\%  &78\%  \\ $P_{2}$ ($\uparrow - \downarrow \over \uparrow +
\downarrow$) & $-$6\%  &84\% \\ \br
\end{tabular}
\end{indented}
\end{table}


\newpage

\begin{thebibliography}{99}
\bibitem{iosif2}
A detailed study of the full-Heusler alloys would be presented
elsewhere.
\end{thebibliography}
\end{document}